\documentclass[aps,prd,twocolumn,amsmath,superscriptaddress,nofootinbib]{revtex4-1}

\usepackage{varwidth}
\usepackage{multirow}
\usepackage{dcolumn}
\usepackage{tabularx}
\usepackage{booktabs}
\newcolumntype{C}{>{\centering\arraybackslash}X}

\usepackage{graphicx}
\usepackage{epstopdf}
\usepackage{graphics}
\usepackage{graphicx}
\usepackage{epsfig}
\usepackage{epstopdf}
\usepackage{makecell}
\usepackage{hyperref}
\hypersetup{
    colorlinks=true,
    linkcolor=blue,
    citecolor=blue,
    urlcolor=blue,
    anchorcolor=blue}

\begin{document}
\title{Confirming the glueball-like particle $\rm X(2370)$ productions in $e^+e^-$ collisions
       \\ at BESIII energy with PACIAE model}

\author{Zhi-Lei She}
\email[]{shezhilei@cug.edu.cn}
\affiliation{School of Mathematical and Physical Sciences, Wuhan Textile
            University, Wuhan 430200, China}

\author{An-Ke Lei}
\affiliation{Key Laboratory of Quark and Lepton Physics (MOE) and Institute of
            Particle Physics, Central China Normal University, Wuhan 430079,
            China}

\author{Wen-Chao Zhang}
\email[]{wenchao.zhang@snnu.edu.cn}
\affiliation{School of Physics and Information Technology, Shaanxi Normal
University, Xi'an 710119, China}

\author{Yu-Liang Yan}
\affiliation{China Institute of Atomic Energy, P. O. Box 275 (10), Beijing
            102413, China}

\author{Dai-Mei Zhou}
\email[]{zhoudm@mail.ccnu.edu.cn}
\affiliation{Key Laboratory of Quark and Lepton Physics (MOE) and Institute of
            Particle Physics, Central China Normal University, Wuhan 430079,
            China}

\author{Hua Zheng}
\affiliation{School of Physics and Information Technology, Shaanxi Normal
University, Xi'an 710119, China}

\author{Ben-Hao Sa}
\email[]{sabhliuym35@qq.con}
\affiliation{Key Laboratory of Quark and Lepton Physics (MOE) and Institute of
            Particle Physics, Central China Normal University, Wuhan 430079,
            China}
\affiliation{China Institute of Atomic Energy, P. O. Box 275 (10), Beijing
            102413, China}

\date{\today}

\begin{abstract}
The parton and hadron cascade model {\footnotesize PACIAE} is employed to
confirm the BESIII newest observation of glueball-like particle $\rm X(2370)$
production in $e^+e^-$ collisions at $\sqrt{s}=4.95\,\mathrm{GeV}$. We
coalesce the $\rm X(2370)$ glueball state with two gluons in the simulated
partonic final state by the Dynamically Constrained Phase-space Coalescence
({\footnotesize DCPC}) model. Alternative configuration of $\rm X(2370)$ molecular state is
recombined in the simulated hadronic final state with $\pi^{+},\pi^{-},K^{+},
K^{-},K_{S}^{0},K_{S}^{0}$ and $\eta'$ by {\footnotesize DCPC} model. The resulted particle
transverse momentum spectrum and rapidity distribution, etc. show a significant
discrepancy between the two states. They are not only serving as criteria to
distinguish the $\rm X(2370)$ glueball state or molecular state, but also
confirming the BESIII observation of glueball-like particle $\rm X(2370)$
productions in $e^+e^-$ collisions.
\end{abstract}

\maketitle

\section{Introduction}
Quantum chromodynamics (QCD) expects the existence of new types of exotic
hadrons, such as glueballs and hybrids with gluonic degrees of freedom.
Glueballs as the bound states of two and more gluons is an important topic
in the hadron physics~\cite{AMSLER200461,KLEMPT20071,CREDE200974}. $J/\psi$
radiative decay is a good experiment for glueballs production, because of
its gluon-rich environment. The observation of the $\rm X(2370)$ with
spin-parity quantum number of $J^{PC}=0^{-+}$~\cite{PhysRevLett.132.181901} is
crucial. Recently it is confirmed by the Lattice QCD (LQCD) calculations and
regarded as the pseudoscalar glueball
candidate~\cite{PhysRevD.73.014516,IJMPE2009,PhysRevD.100.054511}.
The $\rm X(2370)$ has been sequentially certificated in
$J/\psi \rightarrow \gamma \pi^{+}\pi^{-}\eta'$~\cite{PhysRevLett.106.072002},
$J/\psi \rightarrow \gamma K^{+} K^{-} \eta'$ and
$J/\psi \rightarrow \gamma K_{S}^{0} K_{S}^{0} \eta'$~\cite{Ablikim2020}.
Most recently it has been measured again by the BESIII in $e^{+}e^{-}$
collisions at $\sqrt{s}=4.95\,\mathrm{GeV}$ and regards as glueball-like
particle. The measured mass and width of $\rm X(2370)$ is
$2395\pm11(\mathrm {stat})^{+26}_{-94}(\mathrm {syst})\,\mathrm{ MeV/c^{2}}$,
and $188^{+18}_{-17}(\mathrm {stat})^{+124}_{-33}(\mathrm {syst})\,\mathrm{ MeV/c^{2}}$ ~\cite{PhysRevLett.132.181901}, respectively.

Various explanations for the $\rm X(2370)$ have been proposed, such
as a pseudoscalar
glueball~\cite{PhysRevD.100.054511,PhysRevD.87.054036,ZHANG2022136960}, radial
excitation of $\eta'$ meson~\cite{PhysRevD.83.114007,PhysRevD.102.114034},
a mixture state of pseudoscalar glueball and radial
excitation of $\eta'$~\cite{PhysRevD.82.074026},
a $ss \overline{s} \overline{s}$ tetraquark state~\cite{PhysRevD.106.014023},
compact hexaquark states~\cite{PhysRevD.86.014008}, etc. Despite significant
efforts have been made, little consensus on its nature is still lacking
indeed~\cite{chen2023}.

In this paper, we employ the parton and hadron cascade model
{\footnotesize PACIAE}~\cite{lei2023} together with the Dynamically
Constrained Phase space Coalescence ({\footnotesize DCPC})
model~\cite{yan2012} to confirm the BESIII newest observation of
glueball-like particle $\rm X(2370)$ production in $e^+e^-$ collisions at
$\sqrt{s}=4.95\,\mathrm{GeV}$. Based on the {\footnotesize PACIAE} simulated
Partonic Final State (PFS) we coalesce the $\rm X(2370)$ glueball state with
two gluons by {\footnotesize DCPC} model. The $\rm X(2370)$ molecular state is
recombined in the simulated Hadronic Final State (HFS) with
$\pi^{+}+\pi^{-}+\eta'$ or $K^{+}+K^{-}+\eta'$ or $K_{S}^{0}+K_{S}^{0}+\eta'$
by {\footnotesize DCPC} model. Our results of the $\rm X(2370)$ particle
yield, its rapidity ($y$) and transverse momentum ($p_T$) single differential
distributions, as well as its $p_T$ and $y$ double differential distributions
show a significant discrepancy between two states. These observables
effectively distinguish the $\rm X(2370)$ glueball state from $\rm X(2370)$
molecular state and thus confirm the BESIII observation of glueball-like
particle $\rm X(2370)$ production in $e^{+}e^{-}$ collisions.

The {\footnotesize PACIAE + DCPC} model is a phenomenological model for
elementary particle collisions and heavy-ion collisions. Its Monte Carlo (MC)
simulation version has been successfully applied describing the production of
exotic hadrons, such as $\mathrm X(3872)$, $P_{c}$ states, and
$\chi_{c1}(3872)/\psi(2S)$ cross-section ratio,
in relativistic nuclear collisions~\cite{ge2021,hui2022,tai2023,she2024prc}.
In particular, the $\rm X(3872)$ tetraquark state and the molecular state have
been coalesced and recombined, respectively, in the partonic final state and
the hadronic final state~\cite{she2024prc}.

\section{Models}
\subsection{Brief introduction to $e^{+}e^{-}$ collisions in PACIAE}
The {\footnotesize PACIAE} model~\cite{sa2012,zhou201589,lei2023} is based on
the {\footnotesize PYTHIA 6.4}~\cite{sjostrand2006} with additionally
considering the partonic rescattering and the hadronic rescattering before and
after hadronization, respectively. The latest version of
{\footnotesize PACIAE 3.0}~\cite{lei2023} is employed in this work.

In {\footnotesize PACIAE} model each positron-electron ($e^{+}e^{-}$)
collision is executed by {\footnotesize PYTHIA}~\cite{sjostrand2006} with
presetting the hadronization turning-off temporarily. A partonic initial
state is then available. This partonic initial state undergos the partonic
rescattering, where the Lowest-Order perturbative Quantum ChromoDynamics
(LO-pQCD) parton-parton interaction cross section~\cite{combridge} is
employed. The resulted partonic final state comprises a lots and lots of
quarks and (anti)quarks as well as gluons with their four-coordinate
and four-momentum.

\begin{center}
\begin{figure}[htbp]
\centering
\includegraphics[width=0.45\textwidth]{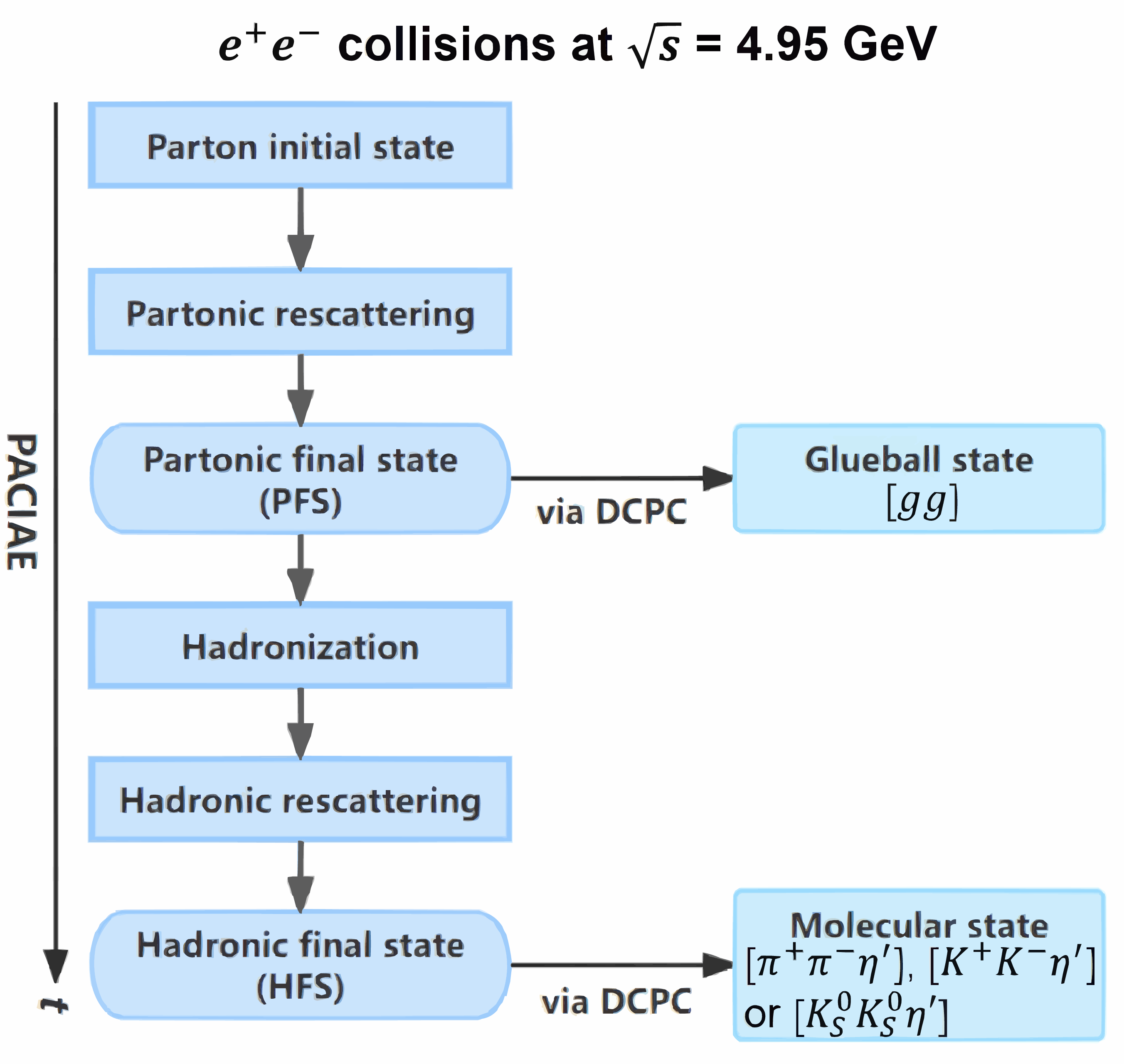}
\caption{A sketch for the $\rm X(2370)$ glueball state and molecular state
productions in $e^{+}e^{-}$ collision at $\sqrt{s}= 4.95\,\mathrm {GeV}$ in
{\footnotesize PACIAE + DCPC} model.}
\label{sket1}
\end{figure}
\end{center}

After partonic rescattering, the hadronization is implemented by the Lund
string fragmentation regime and/or the coalescence model~\cite{lei2023}
generating a hadronic intermediate state. It is followed by the hadronic
rescattering resulted a hadronic final state for a $e^{+}e^{-}$ collision.
This hadronic final state is composed of numerous hadrons with their
four-coordinate and four-momentum. A block diagram of above transport
processes is shown in the left part of the Fig.~\ref{sket1}.

\subsection{DCPC model}
In the quantum statistical mechanics~\cite{stowe2007,kubo1965}, the yield of
a cluster composed of N particle (the parton in PFS or the hadron in HFS) is
\begin{eqnarray}
Y_{N}=\int\cdots\int_{E_\alpha\le E\le E_\beta}\frac{d\vec{q}_{1}d\vec{p}_{1}\cdots d\vec{q}_{N}d\vec{p}_{N}}{h^{3N}},
\label{eq: two}
\end{eqnarray}
where the $E_{\alpha}$ and $E_{\beta }$ are particle lower and upper energy
thresholds. The $\vec {q}_{i}$ and $\vec {p}_{i}$ are, respectively, the
$i$th particle three-coordinate and three-momentum. If the cluster exists
naturally, certain constraints (the component constraint, coordinate
constraint, and the momentum constraint) should be satisfied. Therefore the
yield of two gluons cluster, for instance, is calculated by

\begin{eqnarray}
Y_{gg}=\frac{1}{2!} \int \dots \int {\delta_{12}
	\frac{d\vec{q}_{1}d\vec{p}_{1}d\vec{q}_{2}d\vec{p}_{2}}{h^{6}}}
\label{eq: three},
\end{eqnarray}
where
\begin{eqnarray}
\delta_{12}=\left\{\begin{array}{ll}
	1  \mbox { if } [1 \equiv g,  2 \equiv \overline g, \\
\quad m_{\rm 0}-\Delta m \leq m_{inv}\leq m_{\rm 0}+\Delta m, \\
         \hspace{0.25cm} |\vec{q}_{ij}| \leq R_{0} \hspace{0.25cm}
	(i\neq j;i,j=1,2)]; \\[4pt]
    0  \mbox { otherwise. }
\end{array}\right.
\label{eq: three_1}
\end{eqnarray}

The factor of ${1}/{2!}$~\cite{kubo1965} in Eq.~(\ref{eq: three}) is
considering two gluons are identical particles. In the Eq.~(\ref{eq: three_1}),
the $m_0$ denotes cluster mass and the $\Delta m$ is mass uncertainty (a
parameter). The invariant mass, $m_{inv}$, reads
\begin{eqnarray}
m_{inv}=\sqrt{\bigg(\sum^{2}_{i=1} E_i \bigg)^2-\bigg(\sum^{2}_{i=1}
\vec p_i \bigg)^2},
\label{eq:two_3}
\end{eqnarray}
where $E_i$ ($i=1,2$) is component particle ($g$) energy. The
$R_0$ and $|\vec{q}_{ij}|$ denote, respectively, the cluster radius
(a free parameter) and the relative distances between two component particles
of $i$ and $j$.

To coalesce the $\rm X(2370)$ two-gluon glueball state, we first construct a
component particle list (gluon list) based on the partons list in PFS. A
two-layer cycle over gluons in the list is then built. Each combination in
this cycle, if it satisfies the constraints of Eq.~(\ref{eq: three_1}), counts
as an $\rm X(2370)$ glueball state. The gluon list is then updated by removing
the used gluons resulting a new gluon list. A new two-layer cycle is built,
etc. Repeat these processes until the empty of gluon list or the rest in the
list is less than two. The parameters in Eq.~(\ref{eq: three_1}) are set as:
$m_0=m_{\rm X(2370)}= 2395\,\mathrm{MeV/c^{2}}$ and
$\Delta m=94\,\mathrm{MeV/c^{2}}$ (estimated by the measured half decay width
of $\rm X(2370)$~\cite{PhysRevLett.132.181901}, as well as
$R_{0}<1.0\,\mathrm{fm}$~\cite{hui2021}.

Similarly, the $\rm X(2370)$ molecular state is recombined by
{\footnotesize DCPC} model in HFS with component mesons of
$\pi^{+}, \pi^{-}, K^{+}, K^{-}, K_{S}^{0}, K_{S}^{0}$ and $\eta'$. Thus we
first construct a component meson list based on the hadron list in HFS.
A three-layer cycle over component mesons in the list is then built.
Each combination in the cycle, if it is ($\pi^{+},\pi^{-},\eta'$) or
($K^{+},K^{-},\eta'$) or ($K_{S}^{0},K_{S}^{0},\eta'$) and satisfies the
corresponding constraints similar to Eq.~(\ref{eq: three_1}) but without extra
factor of ${1}/{2!}$, counts as an $\rm X(2370)$ molecular state. Here the
corresponding parameters are set as follows: the $m_{\rm X(2370)}$ and
$\Delta m$ are the same ones in the glueball state calculation. The $R_{0}$
sets in the range of $1.0\,\mathrm{fm} < R_{0} < 3.0\,\mathrm{fm}$,
here $R_{0}$ upper bound is assumed to be a summation of radius of
three component mesons.

\section{Result and discussion}
We generate $4 \times 10^{8}$ $\sqrt{s}=4.95\,\mathrm{GeV}$ $e^{+}e^{-}$
collision events by {\footnotesize PACIAE 3.0} model with default parameters.
The $\rm X(2370)$ glueball state and molecular state are then coalesced and
recombined by~{\footnotesize DCPC} model in the PFS and HFS, respectively,
as shown in Fig.~\ref{sket1}.

\begin{figure*}[!htbp]
\includegraphics[scale=1.08]{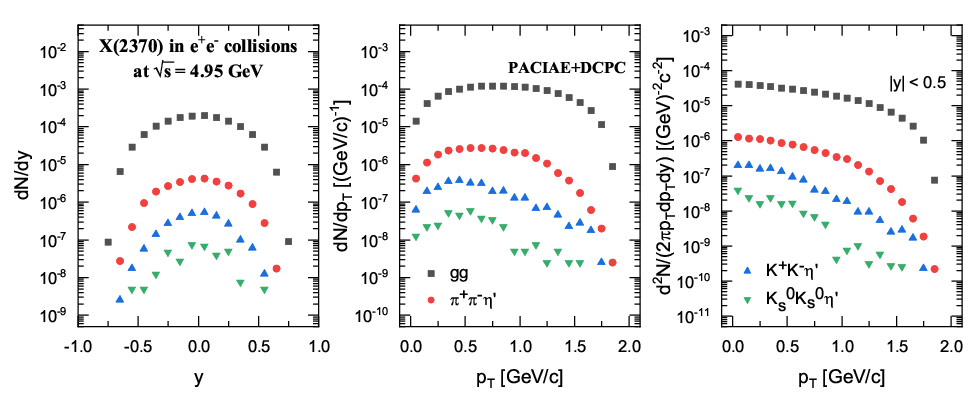}
\caption{The simulated $y$ and $p_T$ single differential distribution, as well
as $p_T$ and $y$ double differential distribution of $\rm X(2370)$ two states
in $e^{+}e^{-}$ collisions at $\sqrt{s}=4.95\,\mathrm {GeV}$.}
\label{common}
\end{figure*}

The particle single $y$ and $p_T$ differential distributions, as well as the
$p_T$ and $y$ double differential distributions of $\rm X(2370)$ two states
are calculated and given in Fig.~\ref{common}. A significant discrepancy
between $\rm X(2370)$ two states is observed in all the above distributions.
They are not only serving as criteria to distinguish the $\rm X(2370)$
glueball state from the molecular state, but also confirming the BESIII
observation of glueball-like particle $\rm X(2370)$ production in $e^+e^-$
collisions.

The yield of the molecular state shows a visible difference among
($\pi^{+}\pi^{-}\eta'$), ($K^{+}K^{-}\eta'$), and ($K_{S}^{0}K_{S}^{0}\eta'$).
$\pi^{+}\pi^{-}\eta'$ is the largest and $K_{S}^{0}K_{S}^{0}\eta'$ is the
least. This is consistent with the order in $\rm X(2370)$ radiative decay
branching fraction~\cite{Ablikim2020,PhysRevD.87.054036}.

We consider only $\rm X(2370)$ two-gluon glueball state because of the BESIII
measured $\rm X(2370)$ mass ~\cite{PhysRevLett.132.181901} is close to the
LQCD prediction for a two-gluon pseudoscalar glueball of
$2.3-2.6\,\mathrm{GeV/c^{2}}$~\cite{PhysRevD.73.014516,IJMPE2009,PhysRevD.100.054511}.
Noteworthy, the predicted three-gluon pseudoscalar glueball
mass is around $3.4-4.5\,\mathrm{GeV/c^{2}}$ in both
the LQCD~\cite{IJMPE2009,PhysRevD.100.054511,jhep201210} and the QCD sum
rule~\cite{PhysRevD.104.094050}.

As for the decay of pseudoscalar glueball, we just consider its
three-pseudoscalar-meson channel of $G \rightarrow \pi^{+}\pi^{-}\eta'$,
$G \rightarrow K^{+}K^{-}\eta'$ and $G \rightarrow K_{S}^{0}K_{S}^{0}\eta'$
(here $G$ refers to the glueball). There is another channel of a light
pseudoscalar meson (such as $\pi, K, \eta')$ and a quark-antiquark nonet of
scalars above 1 GeV (such as $a_0(1450), K_0^*(1430), f_0(1370), f_0(1500)$ or
$f_0(1710)$)~\cite{PhysRevD.87.054036,PhysRevD.100.054511}. Therefore the
$X(2370)$ molecular state recombined by a light pseudoscalar meson and a
quark-antiquark nonet has to be studied in the next work.

The similar study should also be extended to $pp$ collisions and Pb-Pb
collisions at LHC energies. We are also planned investigating the $X(2370)$
nuclear modification factor ($R_{AA}$) and azimuthal asymmetry ($v_2$) as
criterion to distinguish $X(2370)$ glueball state from the molecular state.
Meanwhile, the glueball nature, especially the three-gluon glueball
configuration, has to be studied further. At last, we propose that all kinds
of theoretical models (such as the LQCD, the QCD sum rule, and our
phenomenological model Monte Carlo simulation method, etc.) study together
with the experiments to unravel the nature of the glueball.

\section*{Acknowledgements}
We thank Shan Jin, Yan-Ping Huang, and Li-Lin Zhu for helpful
discussions. This work was supported by the National Natural Science
Foundation of China under grant Nos.: 12375135 and by the 111 project of the
foreign expert bureau of China. Y.L. Yan acknowledges the financial support
from Key Laboratory of Quark and Lepton Physics in Central
China Normal University under grant No. QLPL201805 and the Continuous Basic
Scientific Research Project (No, WDJC-2019-13). W.C. Zhang is supported
by the Natural Science Basic Research Plan in Shaanxi Province of China
(No. 2023-JCYB-012).


%

\end{document}